\documentclass[10pt,journal,compsoc]{IEEEtran}

\usepackage{times}
\usepackage{soul}
\usepackage{url}
\usepackage[hidelinks]{hyperref}
\usepackage[utf8]{inputenc}
\usepackage[small]{caption}
\usepackage{graphicx}
\usepackage{amsmath}
\usepackage{amsthm}
\usepackage{booktabs}
\usepackage{algorithm}
\usepackage{algorithmic}
\usepackage{subfigure}

\usepackage{enumitem}

\usepackage{tabularx}
\usepackage{amsfonts}
\usepackage{multirow}
\usepackage{color}
\usepackage{bbm}

\newcolumntype{L}[1]{>{\raggedright\arraybackslash}p{#1}}
\newcolumntype{C}[1]{>{\centering\arraybackslash}p{#1}}
\newcolumntype{R}[1]{>{\raggedleft\arraybackslash}p{#1}}

\ifCLASSOPTIONcompsoc
  \usepackage[nocompress]{cite}
\else
  \usepackage{cite}
\fi

\ifCLASSINFOpdf
\else
\fi

\begin{document}
\title{Scene-adaptive Knowledge Distillation for Sequential Recommendation via Differentiable Architecture Search~\thanks{Lei Chen and Fajie Yuan contribute equally.}}

\author{Lei~Chen,
        Fajie~Yuan,
        Jiaxi~Yang,
        Min~Yang,
        and~Chengming~Li
\IEEEcompsocitemizethanks{
\IEEEcompsocthanksitem Lei Chen, Min Yang and Chengming Li are with Shenzhen Institute of Advanced Technology, Chinese Academy of Sciences, Shenzhen 518055, P.R. China.\protect
E-mail: lei.chen@siat.ac.cn, min.yang@siat.ac.cn and cm.li@siat.ac.cn.
\IEEEcompsocthanksitem Fajie Yuan is with Westlake University, Hangzhou 310024, P.R. China. A part of this work was finished when Fajie was AI researcher at Tencent Kandian Group.
\protect
E-mail: yuanfajie@westlake.edu.cn or fajieyuan@tencent.com.
\IEEEcompsocthanksitem Jiaxi Yang is with Huazhong University of Science and Technology, Wuhan 430074, P.R. China. \protect
E-mail: yangjiaxi@hust.edu.cn.
}
}


\markboth{MANUSCRIPT}
{Shell \MakeLowercase{\textit{et al.}}: Bare Demo of IEEEtran.cls for Computer Society Journals}

\IEEEtitleabstractindextext{%
\begin{abstract}
Sequential recommender systems (SRS) have become a research hotspot due to its power in modeling user dynamic interests and sequential behavioral patterns. To maximize model expressive ability, a default choice is to apply a larger and deeper network architecture, which, however, often brings high network latency when generating online recommendations. Naturally, we argue that compressing the heavy recommendation models into middle- or light- weight neural networks is of great importance for practical production systems. To realize such a goal, we propose AdaRec, a knowledge distillation (KD) framework which compresses knowledge of a teacher model into a student model adaptively according to its recommendation scene by using differentiable Neural Architecture Search (NAS). Specifically, we introduce a target-oriented distillation loss to guide the structure search process for finding the student network architecture, and an cost-sensitive loss as constraints for model size, which achieves a superior trade-off between recommendation effectiveness and efficiency. In addition, we leverage Earth Mover's Distance (EMD) to realize many-to-many layer mapping during knowledge distillation, which enables each intermediate student layer to learn from other intermediate teacher layers adaptively. Extensive experiments on real-world recommendation datasets demonstrate that our model achieves competitive or better accuracy with notable inference speedup comparing to strong counterparts, while discovering diverse neural architectures for sequential recommender models under different recommendation scenes.
\end{abstract}

}

\maketitle

\IEEEdisplaynontitleabstractindextext

\IEEEpeerreviewmaketitle

\IEEEraisesectionheading{\section{Introduction}\label{sec:introduction}}

\IEEEPARstart{S}{equential}  (a.k.a. session-based) recommender systems that aim to predict new interactions  based on user historical ones have attracted much attention in recent years. Particularly, with the tremendous success of deep learning, deep neural network (DNN) based sequential recommendation (SR) models have yielded substantial improvements comparing to traditional collaborative filtering (CF)~\cite{hidasi2015session}, such as neighborhood methods~\cite{sarwar2001item} and shallow factorization models~\cite{koren2009matrix}. This is because with many hidden layers, well-designed  deep models could be more powerful in capturing user dynamic interests, high-level or long-range sequential relations of user interactions. More recently, Wang \textit{et al.} \cite{wang2020stackrec} and Chen \textit{et al.} \cite{chen2021user} revealed that deep SR models
such as NextItNet~\cite{yuan2019simple} and SASRec~\cite{kang2018self} could be stacked in a surprised depth with over 100 layers for achieving their optimal performance.

However, a real problem arises as these deep SR models go bigger and deeper; that is, the model becomes too large in parameter size, and both memory and inference costs increase sharply,  making the deployment of them difficult in production systems.
Thereby, we argue that compressing the heavy deep SR models into moderate- or light-weight neural networks without sacrificing their accuracy is of crucial importance for practical usage. Knowledge Distillation  (KD)~\cite{hinton2015distilling} as an effective compression technique has been recently investigated in the recommender systems domain~\cite{tang2018ranking,pan2019novel}. By  transferring 
useful knowledge from a big teacher network to the student network, large deep models could be slimmed into a smaller and shallower structure without performance degradation.
However, existing KD methods basically distill the teacher model into a fixed-structure student model that is manually designed in advance.
This potentially limits the flexibility and scalability of the student model, especially for diverse and relatively complicated scenarios in recommender systems. For example, the optimal structure for music recommendation might be different from the optimal structure for E-commerce recommendation. 
 Ideally, we hope to generate an adaptive student model whose optimal structure  considers the specific recommendation scenarios. 
 
Inspired by the success of automated machine learning (AutoML), we propose a novel knowledge distillation method to compress the deep sequential recommendation models, termed AdaRec. AdaRec distills the knowledge of a teacher model into a student model adaptively according to the recommendation scene based on differentiable Neural Architecture Search (NAS)~\cite{liu2018darts,xie2018snas,chen2020adabert}. Specifically, we devise a target-oriented knowledge distillation loss to provide search 
supervision for learning the architecture of student network, and a cost-sensitive loss as additional regularization to constrain the model size, which achieve a superior trade-off between recommendation effectiveness and efficiency. In addition, we leverage Earth Mover's Distance (EMD) to realize  effective many-to-many layer mapping during the distillation process,  enabling each intermediate  layer of student  to learn from any other intermediate  layers of its teacher. 
It is worth noting that, our method is a generic knowledge distillation framework which can directly apply to a broad class of well-known sequential recommendation models, such as NextItNet~\cite{yuan2019simple}, SASRec~\cite{kang2018self} and BERT4Rec~\cite{sun2019bert4rec}.
Besides, with the well-designed NAS architecture, our method  can distill the deep sequential recommendation models into effective smaller models with diverse neural network architectures, according to the specific recommendation scenarios.

 Our contributions in this paper are fourfold:
\begin{itemize}
    \item To the best of our knowledge, we are the first to propose combining Knowledge Distillation and Neural Architecture Search in the SRS tasks so as to adaptively compress the deep sequential recommendation models according to recommendation scenes.
    \item We devise a knowledge distillation loss  based on Earth Mover's Distance (EMD) and a cost-sensitive constraint to achieve a trade-off between recommendation effectiveness and efficiency. 
    \item We verify the universality of the AdaRec framework by performing KD with three different teacher models, namely, NextItNet~\cite{yuan2019simple}, SASRec~\cite{kang2018self} and BERT4Rec~\cite{sun2019bert4rec}.
    \item We conduct extensive experiments on three real-world recommendation datasets with different scenarios, demonstrating that AdaRec achieves competitive or better accuracy with notable inference speedup comparing to its original teacher model. Moreover, we discover diverse neural architectures of the student model in different recommendation scenarios or tasks.  
\end{itemize}

\section{Related Work}
\subsection{Deep Sequential Recommendation}
Sequential (a.k.a. session-based) recommender systems (SRS) is an important branch in the recommendation field and has become a hotspot recently due to the wide range of application scenarios and huge commercial values. Since in this paper we focus on compressing large and deep sequential recommendation models, we only review related work regarding its advancement in deep learning (DL).


Deep neural networks have achieved superior recommendation accuracy in SRS tasks. 
In general, these models could be classified into three categories, namely RNN-based, CNN-based and self-attention based methods. 
 Specifically, Hidasi \textit{et al.} \cite{hidasi2015session} proposed GRU4Rec, which is the first RNN-based sequential recommendation model. Following this work, many extended works were proposed, which either optimized a new ranking loss~\cite{hidasi2018recurrent}, incorporated more context features~\cite{gabriel2019contextual}, or developed more advanced data augmentation~\cite{tan2016improved}. 
 While effective, these models rely heavily on the hidden states of the entire past, which cannot take full advantage of the parallel processing resources (e.g., GPU and TPU)~\cite{yuan2019simple} during training. Therefore, Convolutional Neural Network (CNN) and self-attention based models are proposed to mitigate such limitations~\cite{tang2018personalized,yuan2019simple,kang2018self,sun2019bert4rec}. 
 Among them, Tang \textit{et al.} \cite{tang2018personalized} proposed Caser,
  which embeds a sequence of user-item interactions into an ``image'' and learn sequential patterns as local features of the image by using wide convolutional filters. Subsequently,
\cite{yuan2019simple} proposed NextItNet, a very deep 1D temporal CNN-based  recommendation model which particularly excels at modeling long-range item sequences.
In addition, self-attention based  models, such as SASRec~\cite{kang2018self} and BERT4Rec~\cite{sun2019bert4rec}, also showed competitive accuracy for SRS tasks as well. 
SASRec~\cite{kang2018self} utilized the popular self-attention mechanism to model long-term sequential semantics by encoding user’s historical behaviors.
Inspired by the great success of BERT~\cite{devlin2018bert} in NLP filed, Sun \textit{et al.} \cite{sun2019bert4rec} proposed BERT4Rec, which uses the transformer architecture and masked language model to learn bidirectional item dependencies for better sequential recommendations. In this paper, we present AdaRec by applying NextItNet, SASRec \& BERT4Rec as
 teacher networks  given their superior performance in literature.
In addition, Wu \textit{et al.} \cite{wu2020sse} proposed SSE-PT, a personalized transformer model which applies stochastic shared embeddings (SSE) regularization to achieve personalized user representations.
With the advancement on Graph Neural Networks (GNN), GNN-based sequential  models, such as SR-GNN~\cite{wu2019session}, GC-SAN~\cite{xu2019graph} and MA-SAN~\cite{ma2020memory}, have also attracted attention
yielded substantial improvements in recommendation accuracy. 
Besides, there are some other works~\cite{ma2019hierarchical,lv2019sdm,peng2020ham} that designs novel neural network modules for sequential recommendations. For instance, HGN~\cite{ma2019hierarchical}, a hierarchical gating neural network, adopts a feature gating and an instance gating to determine what item features should be used for recommendation. SDM~\cite{lv2019sdm} integrates a multi-head self-attention module with a gated fusion module to capture both short- and long-term user preferences. HAM~\cite{peng2020ham} develops hybrid associations models to further capture sequential and multi-order user-item association patterns for sequential recommendations.

\subsection{Knowledge Distillation}
Large and deep neural networks have achieved remarkable success in recent recommendation literature~\cite{sun2020generic,wang2020stackrec,sanh2019distilbert}. However, the deployment of such heavy model for real production system remains a great challenge. Knowledge Distillation (KD)~\cite{hinton2015distilling,li2020bert} is a representative technique for model compression and acceleration. Its basic idea is to transfer important knowledge from a big teacher network to a small student network. Specifically, Tang \textit{et al.} \cite{tang2018ranking} proposed the first
KD technique for learning to rank problems in recommender systems. However, the work only focused on distillation on very shallow neural recommendation models while its effectiveness for deep SRS keeps largely unknown. ~\cite{liu2020general} presented a general knowledge distillation framework for counterfactual recommendation with four types of distillation, namely, label-based, feature-based, sample-based and model structure-based distillation. More recently,~\cite{kang2020rrd} proposed a  knowledge distillation framework that
forces the student network to learn  from both the teacher’s output and the latent knowledge stored in the teacher model. 

KD-based compression have also been widely studied in other domains~\cite{sun2019patient,sanh2019distilbert,jiao2019tinybert,li2020bert}. Nowak \textit{et al.} \cite{nowak2018deep} proposed a structure compression method which involves transferring the knowledge learned by multiple layers to a single layer. Wang \textit{et al.} \cite{wang2018progressive} progressively performed block-wise knowledge transfer from teacher networks to student networks while preserving the receptive field. Mirzadeh \textit{et al.} \cite{mirzadeh2020improved} introduced a teacher assistant to mitigate the training gap between teacher model and student model. Recently, compressing pretrained language models (e.g., BERT) with KD has attracted increasing attention, and numerous novel models are proposed to effectively distill BERT from different perspectives (e.g., embedding layer, hidden layers and prediction layer), such as PKD-BERT~\cite{sun2019patient}, DistilBERT~\cite{sanh2019distilbert}, TinyBERT~\cite{jiao2019tinybert} and BERT-EMD~\cite{li2020bert}. 

\begin{figure*}[htbp]
    \centering
    \includegraphics[scale=0.9]{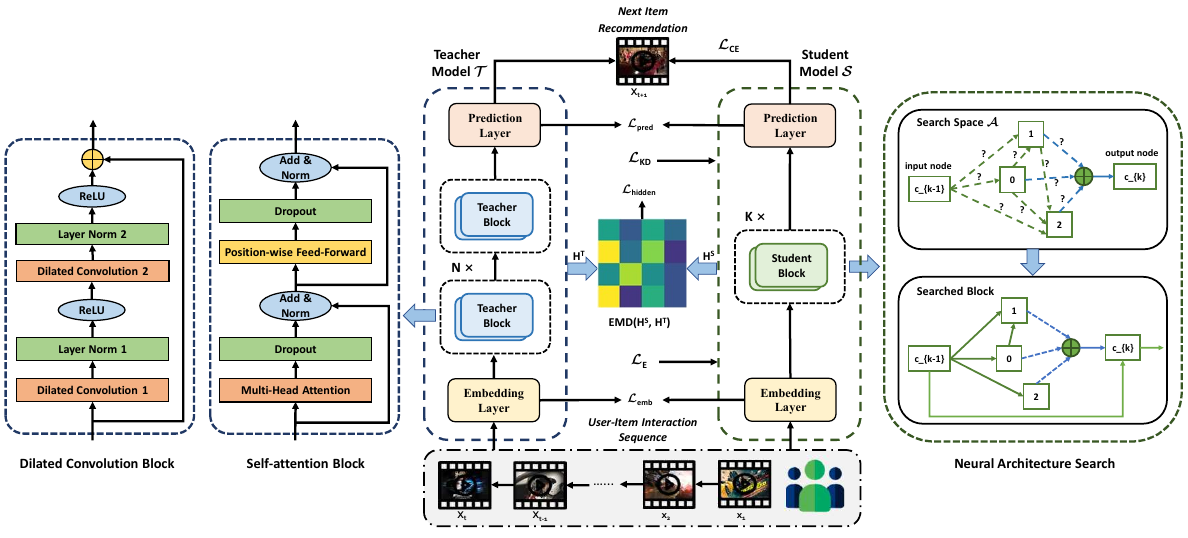}
    \caption{Model architecture of AdaRec. The proposed AdaRec consists of two primary components: teacher model and student model. In this paper, we specify AdaRec using NextItNet~\cite{yuan2019simple}, SASRec~\cite{kang2018self} and BERT4Rec~\cite{sun2019bert4rec} as the teacher models given their superior recommendation performance. The structures of the student models are searched based on neural architecture search techniques in a differentiable manner. Specifically, we devise a target-oriented knowledge distillation loss to provide search hints for searching the architecture of student network, and an efficiency-aware loss as search constraints for constraining the model size, which achieves a superior trade-off between effectiveness and efficiency for sequential recommendations.}
    \label{fig:model}
\end{figure*}

\subsection{Neural Architecture Search}
Neural Architecture Search (NAS) that automatically discovers the  network architecture, has gained increasing attention recently. Early NAS methods based on reinforcement learning \cite{zoph2016neural} and evolution \cite{real2019regularized} are computationally very expensive. Recent studies significantly speed up the search and evaluation stages by architecture parameter sharing, such as ENAS \cite{pham2018efficient}, gradient-descent based DARTS \cite{liu2018darts,chen2020adabert} and SNAS \cite{xie2018snas}, and hardware-aware optimization such as AMC \cite{he2018amc} and FBNet \cite{wu2019fbnet,wan2020fbnetv2}.
Different from existing work, we devise a target-oriented knowledge distillation loss to provide search supervision for learning the architecture of the student network, which is a joint search of student structure and knowledge transfer under the guidance of the teacher model. To our best knowledge, we are the first to propose a combination of KD and NAS for compressing the deep sequential recommendation models.

\section{Problem Definition}
Given a sequence of user’s historical behaviors $X^u=[x_1^u,x_2^u,\ldots, x_{t}^u]$ (interchangeably denoted by $x_{1:t}^u$), where $x_{t}^u$ denotes the $t$-th interacted item of user $u$, the goal of SRS is to infer the item $x_{t+1}^u$ that the user would like to interact with at time $t+1$. Since  users usually pay attention to only the first few items, the top-$N$ items are recommended, referred to as the top-$N$ item recommendation problem. 

Deep neural networks, such as NextItNet \cite{yuan2019simple}, SASRec \cite{kang2018self}, BERT4Rec \cite{sun2019bert4rec}, have been proposed and deployed to many SRS applications, yielding state-of-the-art performances. However, different from many traditional collaborative filtering models~\cite{he2017neural}, deep sequential recommendation models often require  more hidden layers to model complex and long-term relations of user actions. Recent work in ~\cite{sun2020generic} showed that the state-of-the-art temporal CNN model
NextItNet needs over 30 layers to reach its maximum expressive ability on some benchmark datasets. Even more, ~\cite{wang2020stackrec,chen2021user} demonstrated by experiments that both NextItNet
and SASRec~\cite{kang2018self} should be stacked with over 100 layers for achieving their best accuracy. This could lead to a large model size and high network latency in practice,
 bringing difficulties for the deployment of them in production systems. 
Therefore, in this paper, we hope to  reduce the model size and accelerate the inference speed for these very deep sequential recommendation models without sacrificing their  accuracy. 

\section{AdaRec}
We introduce a novel scene-adaptive KD-based model compression approach with differentiable NAS, called AdaRec.  
Formally, 
suppose that a large teacher model $\mathcal{T}$ is trained on a target dataset $D$, and the architecture searching space is denoted as $\mathcal{A}$. The goal of AdaRec is to automatically find a high-performing student model $\mathcal{S}$ from  $\mathcal{A}$ with a smaller scale. 
Figure \ref{fig:model} illustrates the overview of the AdaRec framework.
The basic idea is to transfer knowledge from a large teacher recommender model $\mathcal{T}$ to a small student model $\mathcal{S}$ adaptively subject to the specific recommendation task.
In this paper, we specify AdaRec using NextItNet~\cite{yuan2019simple}, SASRec~\cite{kang2018self} and BERT4Rec~\cite{sun2019bert4rec} learning algorithm as the teacher models given their superior recommendation performance. It is noteworthy that the ``teacher'' is model-agnostic and potentially applicable for any sequential recommendation model with a deep network architecture. 
Specifically, the network structures of the student model are automatically searched based on the NAS techniques. To this end, we devise a KD loss to provide search supervision for learning the architecture of the student network and a cost-sensitive loss as search regularization to control the model size. In this manner, our AdaRec could achieve a superior trade-off between effectiveness and efficiency for SRS tasks.

In what follows, we describe AdaRec by elaborating its teacher model, student model, the KD process and the NAS searching process.

\subsection{Teacher Model}
\label{teacher}
 We employ the block-wise (e.g., ResNet~\cite{he2016deep}) deep networks as the teacher models given their powerful performance in literature.
The general framework of the teacher model consists of the bottom embedding layer, hidden layers and the softmax layer. 

In terms of the hidden layers,  we use the residual blocks from  NextItNet~\cite{yuan2019simple}, SASRec~\cite{kang2018self} and BERT4Rec~\cite{sun2019bert4rec} for case study, where  NextItNet is based on the dilated CNN blocks, SASRec
and BERT4Rec are based on the self-attention blocks. The residual block structures are depicted in Figure~\ref{fig:model}.

\paragraph{NextItNet} NextItNet is composed of a stack of dilated convolutional (DC) layers, which are wrapped by a residual block structure every two layers. Specifically, each input item $x^u$ is  converted into an embedding vector $\mathbf{e}^u$, and the user-item interaction sequence $X^u$ is thereby represented by an embedding matrix $\mathbf{E}^u= [\mathbf{e}_1^u \ldots \mathbf{e}_t^u]$. The embedding sequence $\mathbf{E}^u$ is then passed into a stack of dilated convolutional layers to learn feature vector $\mathbf{E}^u_{l}$ which is expected to capture the long-range dependencies. Here, $l$ represents the $l$-th residual block and each residual block  connects two consecutive DC layers. Formally, the $l$-th residual block with the DC operation is formalized as:
\begin{equation}
\mathbf{E}^u_{l} =  \lambda\times\mathcal{F}_{l}(\mathbf{E}^u_{l-1}) + \mathbf{E}^u_{l-1}
\end{equation}
where $\mathbf{E}^u_{l-1}$ and $\mathbf{E}^u_{l}$ are input and output of the $l$-th residual block considered. $\lambda\times\mathcal{F}_{l}(\mathbf{E}^u_{l-1}) + \mathbf{E}^u_{l-1}$ is a shortcut connection by element-wise addition. Similar to~\cite{wang2020stackrec,chen2021user,xiao2018dynamical}, we add a learnable coefficient $\lambda$ to the residual mappings $\mathcal{F}_{l}(\mathbf{E}^u_{l-1})$, so that the model can stack more layers, and get better results than the standard version with $\lambda$ as 1.\footnote{Regarding the effects of the $\lambda$ design, we refer interested users to ~\cite{wang2020stackrec,xiao2018dynamical,bachlechner2020rezero} for detailed analysis.} 
$\mathcal{F}_{l}(\mathbf{E}^u_{l-1})$ represents the residual mapping, which is defined as:
\begin{equation}
\mathcal{F}_{l}(\mathbf{E}^u_{l-1}) =\sigma\left(\mathbf{L} \mathbf{N}_{2}\left(\psi_{2}\left(\sigma\left(\mathbf{L} \mathbf{N}_{1}\left(\psi_{1}(\mathbf{E}^u_{l-1})\right)\right)\right)\right)\right)
\end{equation}
where $\psi_1$ and $\psi_2$ represent the casual convolution operations. $\mathbf{L} \mathbf{N_1}$  and $\mathbf{L} \mathbf{N_2}$ represent layer normalization functions.  $\sigma$ is the ReLU activation function. 

Finally, a softmax output layer is applied to predict the probability distribution for the next item $x^u_{t+1}$: 
\begin{equation}
p(x^u_{t+1}|x^u_{1:t}) = {\rm softmax}(\mathbf{W} \mathbf{E}^u_{l} + \mathbf{b})
\end{equation}
where $\mathbf{W}$ is a projection matrix, and $\mathbf{b}$ is a bias term.

\paragraph{SASRec} Similar to NextItNet, SASRec is composed of a stack of self-attention (SA) layers, which are wrapped by a residual block with a self-attention layer and a feed-forward network. Formally, the $l$-th residual block with the SA operation is formalized as:
\begin{equation}
\label{sasrecfun}
\mathbf{E}^u_{l} =  \lambda\times\mathcal{H}_{l}(\mathbf{E}^u_{l-1}) + \mathbf{E}^u_{l-1}
\end{equation}
where $\mathbf{E}^u_{l-1}$ and $\mathbf{E}^u_{l}$ are input and output of the $l$-th residual block considered. $\lambda\times\mathcal{H}_{l}(\mathbf{E}^u_{l-1}) + \mathbf{E}^u_{l-1}$ is a shortcut connection by element-wise addition. As mentioned above, We also add a learnable coefficient $\lambda$ to the residual mappings $\mathcal{H}_{l}(\mathbf{E}^u_{l-1})$. 
$\mathcal{H}_{l}(\mathbf{E}^u_{l-1})$ represents the residual mapping, which is defined as:
\begin{equation}
\label{sasrecblock}
\mathcal{H}_{l}(\mathbf{E}^u_{l-1}) = \mathbf{\delta}(\mathbf{SA}(\mathbf{L} \mathbf{N_2}(\mathbf{\delta}(\mathbf{FFN}(\mathbf{L} \mathbf{N_1}(\mathbf{E}^u_{l-1}))))))
\end{equation}
where $\mathbf{FFN}$ and $\mathbf{SA}$ represent the feed-forward and self-attention operation, respectively. $\mathbf{L} \mathbf{N_1}$  and $\mathbf{L} \mathbf{N_2}$ represent layer normalization functions.  $\mathbf{\delta}$ is the dropout function. 

Finally, a softmax output layer is applied to predict the probability distribution for the next item $x^u_{t+1}$. For both NextItNet+ and SASRec+, the joint probability $p\left(X^{u}; \Omega\right)$ of each user-item interaction sequence is computed by the product of conditional distributions over interacted items as follows:
\begin{equation}
p\left(X^{u}; \Omega\right) =\prod_{i=2}^{t} p\left(x^u_{i} | x^u_{1: i-1};\Omega \right) p\left(x^u_1\right)
\end{equation}
where $p\left(x_{i}^{u} | x_{1:i-1}^{u}; \Omega\right)$ is the predicted probability for the $i$-th item $x_{i}^{u}$ conditioned on all its previous interactions $[x_{1}^{u}, \ldots, x_{i-1}^{u}]$, and $\Omega$ is the set of parameters.

\paragraph{BERT4Rec} BERT~\cite{devlin2018bert} has shown superior performance in many NLP tasks and the recently proposed BERT4Rec~\cite{sun2019bert4rec} has successfully applied the bidirectional transformer structure to SRS tasks,  which achieves state-of-the-art performance on sequential recommendations. Compared to SASRec  using a unidirectional (left-to-right) transformer structure to capture user’s  dynamic interests, BERT4Rec takes bidirectional dependencies of user’s sequential behaviors into consideration, and proposes a novel Masked Language Model objective to predict the masked items in the interaction sequence. Akin to SASRec, the hidden representations of $l$-th layer in BERT4Rec are defined similarly as  Eq. (\ref{sasrecfun}) and  (\ref{sasrecblock}).

During training, BERT4Rec allows $m$ interactions in the sequence (termed as $x_{\Delta}=[x_{\Delta_{1}}, \ldots, x_{\Delta_{m}}]$) to be masked (i.e., replaced with a special token ``$[mask]$'') and the original interaction sequence $X^{u}$ is modified to $\tilde{X}^{u}$. The goal of the Masked Language Model objective is to predict the original ids of the masked items based solely on its left and right context, which can be formalized as:
\begin{equation}
p\left(X^{u}; \Theta\right) = \prod_{i=1}^{m} p\left(x_{\Delta_{i}} \mid \tilde{X}^{u}; \Theta\right)
\end{equation}
where $\Theta$ is the set of parameters.

Obviously, there is a mismatch between the training and the inference since the Masked Language Model objective aims to predict the current masked items while SRS task aims to predict the next item in the future. To address this, BERT4Rec propose to append the special token ``$[mask]$'' to the end of user’s behavior interaction sequence, and then predict the next item based on the final hidden representation of this token.

\begin{table*}[t]
\centering
\caption{Statistics of the three datasets (after pre-processing). }
    {\renewcommand{\arraystretch}{1.2}
	\centering
 	\small
    \begin{tabular}{L{2.0cm}|C{1.7cm}|C{1.7cm}|C{1.7cm}|C{1.7cm}|C{1.7cm}}
		\toprule
		\textbf{Dataset} & \#Users & \#Items & \#Interactions & \#Sequences & Length $t$  \\ \midrule
		RetailRocket  & 104,593 & 70,012  &  916,421    & 134,241 & 10  \\
		30Music       & 27,364  & 138,990 &  2,081,086  & 177,818 & 20  \\
		ML-2K         & 2,112   & 7,871   &  678,935    & 14,518   & 50 \\
		\bottomrule
	\end{tabular}}
	\label{tab:data_statistic}
\end{table*}

\subsection{Student Model}
Typical model compression methods usually apply KD to transfer knowledge from the large teacher model to the manually designed student model, which rely heavily on the prior knowledge of  human experts to design the structure of the student model.
We perform architecture search of the student network using NAS techniques rather than assigning a fixed  structure in advance. 
Here, we introduce a block-based micro search strategy~\cite{pham2018efficient} to find an optimal network architecture from the search space formed by the operation sets.

\paragraph{Search Space}
The search space design is key to the final performance of the searched student model. In this study, we modularize the large search space of NAS into blocks to reduce its complexity, similar to~\cite{li2020block}.
 In this way, one merely needs to search a few block structures and then repeatedly stack such blocks to form the final network architecture. 
This strategy avoids training each block from scratch, but forces all blocks to share structures, thereby greatly reducing the time to obtain the best performing student model from a large number of candidate networks.
Specifically, the searched block  denoted by $\alpha_{c}$ is represented as a directed acyclic graph (DAG). Each node of the block indicates a latent state $h$ and the edge from node $i$ to  $j$ indicates operation $o_{i,j}$  that transforms $h_i$ to $h_j$. For the $k$-th ($k>1$) searched block, we define an input node $c_{k-1}$, 
 and an output node $c_{k}$ that is obtained by attentively summarized over all intermediate nodes. Formally, let $\mathcal{O}$ be the set of candidate operations, 
 and we assume there is a topological order among $M$ intermediate nodes, i.e., $o_{i,j}\in\mathcal{O}$ exists when $i<j$ and $j\geq1$, the search space $\mathcal{A}$ is thus formalized as:
\begin{equation}
    \mathcal{A}=\alpha_{c}=\left[o_{0,1}, o_{0,2}, o_{1,2}, \ldots, o_{i, j}, \ldots, o_{M, M+1}\right]
\end{equation}

\paragraph{Operation Set} For all the three teacher models (NextItNet, SASRec, BERT4Rec), we adopt the same operation set to search  the student network architecture. 
In this paper, we employ lightweight CNN-based operations as candidates given that they have shown both competitive accuracy and superior efficiency in the SRS literature, compared to RNN~\cite{hidasi2015session} and self-attention~\cite{kang2018self} based models. 
The candidate operation set $\mathcal{O}$ is composed of four kinds of operations: \textit{convolution}, \textit{pooling}, \textit{skip connection}  and \textit{zero} operation.  The \textit{convolution} operations include the 1D convolution, standard
convolutions (without dilation),
 casual dilated convolutions~\cite{yuan2020future} with kernel size $\{3, 5\}$. Note that the dilated convolution is used to capture long-term dependency information. The \textit{pooling} operations include max pooling and average pooling with kernel size $3$. The \textit{skip} operation is leveraged to construct residual connections. The \textit{zero} operation helps to forget the past knowledge.

\begin{table*}[ht]
    \centering
    \caption{Overall performance comparison on the three datasets in terms of MRR@$N$, HR@$N$ and NDCG@$N$ ($N$ is set to 5 and 20). Note that the improvements of AdaRec over all baseline models are statistically significant in terms of paired t-test with p-value $<$ 0.01.}
    \subtable[RetailRocket]{{\renewcommand{\arraystretch}{1.2}
    \begin{tabular}{L{2.5cm}|C{1.26cm}|C{1.26cm}|C{1.26cm}|C{1.26cm}|C{1.26cm}|C{1.26cm}|C{1.26cm}|C{1.26cm}}
    \toprule
    \multirow{2}{*}{\textbf{Model}} & \multicolumn{8}{c}{\textbf{RetailRocket}}\\ \cline{2-9}
    & MRR@5 & MRR@20 & HR@5 & HR@20 & NDCG@5 & NDCG@20 & Params & Speedup \\
    \midrule
    GRU4Rec & 0.6952 & 0.7047 & 0.7748 & 0.8682 & 0.7151 & 0.7421 & $\backslash$ & $\backslash$ \\
    Caser & 0.6489 & 0.6586 & 0.7132 & 0.8106 & 0.6649 & 0.6928 & $\backslash$ & $\backslash$ \\
    \hline
    NextItNet & 0.7139 & 0.7222 & 0.7817 & 0.8645 & 0.7309 & 0.7547 & 40.28M & 1.00$\times$ \\
    KD-NextItNet & 0.7124 & 0.7207 & 0.7889 & 0.8707 & 0.7316 & 0.7552 &  8.80M & 1.97$\times$ \\
    \textbf{AdaRec-NextItNet} &\textbf{0.7345} & \textbf{0.7424} & \textbf{0.7964} & \textbf{0.8741} & \textbf{0.7500} & \textbf{0.7724} &    \textbf{8.66M} & \textbf{2.31$\times$}\\ \hline
    SASRec & 0.6982 & 0.7061 & 0.7511 & 0.8318 & 0.7114 & 0.7343 & 17.80M & 1.00$\times$\\
    KD-SASRec & 0.7221 & 0.7295 & 0.7782 & 0.8525 & 0.7362 & 0.7573 & 4.36M & 2.32$\times$\\
    \textbf{AdaRec-SASRec} & \textbf{0.7352} & \textbf{0.7426} & \textbf{0.7931} & \textbf{0.8682} & \textbf{0.7496} & \textbf{0.7711} &    \textbf{4.34M} & \textbf{6.59$\times$}\\
    \hline
    BERT4Rec & \textbf{0.7561} & 0.7630 & \textbf{0.8150} & \textbf{0.8842} & 0.7709 & \textbf{0.7907} & 18.67M & 1.00$\times$\\
    KD-BERT4Rec & 0.6994 & 0.7085 & 0.7799 & 0.8686 & 0.7196 & 0.7452 & 4.42M & 1.92$\times$ \\
    \textbf{AdaRec-BERT4Rec} & 0.7575 & \textbf{0.7639} & 0.8128 & 0.8759 & \textbf{0.7714} & 0.7895  & \textbf{4.41M} & \textbf{2.07$\times$}\\
    \bottomrule
    \end{tabular}}}
    \subtable[30Music]{{\renewcommand{\arraystretch}{1.2}
    \begin{tabular}{L{2.5cm}|C{1.26cm}|C{1.26cm}|C{1.26cm}|C{1.26cm}|C{1.26cm}|C{1.26cm}|C{1.26cm}|C{1.26cm}}
    \toprule
    \multirow{2}{*}{\textbf{Model}} & \multicolumn{8}{c}{\textbf{30Music}}\\ \cline{2-9}
    & MRR@5 & MRR@20 & HR@5 & HR@20 & NDCG@5 & NDCG@20 & Params & Speedup \\
    \midrule
    GRU4Rec & 0.5242 & 0.5415 & 0.6438 & 0.8133 & 0.5540 & 0.6029 & $\backslash$ & $\backslash$ \\
    Caser & 0.5686 & 0.5787 & 0.6312 & 0.7352 & 0.5842 & 0.6137 & $\backslash$ & $\backslash$ \\
    \hline
    NextItNet & 0.6149 & 0.6282 & 0.7029 & 0.8359 & 0.6368 & 0.6750 & 74.02M & 1.00$\times$ \\
    KD-NextItNet & 0.5969 & 0.6115 & 0.6961 & 0.8402 & 0.6216 & 0.6631 & 17.29M & 1.87$\times$ \\
    \textbf{AdaRec-NextItNet} & \textbf{0.6343} & \textbf{0.6473} & \textbf{0.7151} & \textbf{0.8452} & \textbf{0.6544} & \textbf{0.6917} &  \textbf{17.15M} & \textbf{2.61$\times$} \\ \hline
    SASRec & 0.5761 & 0.5883 & 0.6437 & 0.7692 & 0.5929 & 0.6285 &  34.70M & 1.00$\times$ \\
    KD-SASRec & 0.5881 & 0.6013 & 0.6698 & 0.8033 & 0.6084 & 0.6466 &  8.64M & 2.30$\times$\\
    \textbf{AdaRec-SASRec} & \textbf{0.6132} & \textbf{0.6259} & \textbf{0.6925} & \textbf{0.8321} & \textbf{0.6368} & \textbf{0.6727} &   \textbf{8.62M} & \textbf{5.17$\times$} \\
    \hline
    BERT4Rec & 0.6124 & 0.6249 & 0.7016 & 0.8253 & 0.6347 & 0.6702 &  35.58M & 1.00$\times$ \\
    KD-BERT4Rec & 0.5712 & 0.5881 & 0.6567 & 0.7896 & 0.5820 & 0.6297 &  8.74M & 1.64$\times$ \\
    \textbf{AdaRec-BERT4Rec} & \textbf{0.6262} & \textbf{0.6381} & \textbf{0.7164} &  \textbf{0.8339} & \textbf{0.6487} & \textbf{0.6826} &  \textbf{8.69M} & \textbf{1.84$\times$} \\
    \bottomrule
    \end{tabular}}}
    \subtable[ML-2K]{{\renewcommand{\arraystretch}{1.2}
    \begin{tabular}{L{2.5cm}|C{1.26cm}|C{1.26cm}|C{1.26cm}|C{1.26cm}|C{1.26cm}|C{1.26cm}|C{1.26cm}|C{1.26cm}}
    \toprule
    \multirow{2}{*}{\textbf{Model}} & \multicolumn{8}{c}{\textbf{ML-2K}}\\ \cline{2-9}
    & MRR@5 & MRR@20 & HR@5 & HR@20 & NDCG@5 & NDCG@20 & Params & Speedup \\
    \midrule
    GRU4Rec & 0.4115 & 0.4379 & 0.6141 & 0.8669 & 0.4618 & 0.5355 & $\backslash$ & $\backslash$ \\
    Caser & 0.4186 & 0.4439 & 0.6072 & 0.8465 & 0.4656 & 0.5356 & $\backslash$ & $\backslash$ \\
    \hline
    NextItNet & 0.4453 & 0.4704 & 0.6462 & \textbf{0.8830} & 0.4953 & 0.5648 & 9.87M & 1.00$\times$ \\
    KD-NextItNet & 0.4333 & 0.4584 & 0.6388  & 0.8781 & 0.4844 & 0.5543 & 1.16M & 2.20$\times$  \\
    \textbf{AdaRec-NextItNet} & \textbf{0.4489}  & \textbf{0.4732} & \textbf{0.6519} & 0.8825 & \textbf{0.4995} & \textbf{0.5670} &  \textbf{1.11M} & \textbf{2.78}$\times$ \\ \hline
    SASRec & 0.4241 & 0.4495 & 0.6236 & 0.8654 & 0.4737 & 0.5444 & 2.57M & 1.00$\times$ \\
    KD-SASRec & 0.4137 & 0.4405 & 0.6174 & 0.8719 & 0.4644 & 0.5387 & 0.54M & 2.79$\times$ \\
    \textbf{AdaRec-SASRec} & \textbf{0.4426} & \textbf{0.4669} & \textbf{0.6470} & \textbf{0.8778} & \textbf{0.4934} & \textbf{0.5608} &  \textbf{0.52M} & \textbf{3.81$\times$}\\
    \hline
    BERT4Rec & \textbf{0.4418} & \textbf{0.4667}  & \textbf{0.6502} & \textbf{0.8871} & \textbf{0.4937} & \textbf{0.5629} & 3.45M & 1.00$\times$ \\
    KD-BERT4Rec & 0.4216 & 0.4485 & 0.6269 & 0.8802 & 0.4727 & 0.5470 & 0.54M & 1.58$\times$ \\
    \textbf{AdaRec-BERT4Rec} & 0.4382 & 0.4634  & 0.6471 & 0.8841 & 0.4886 & 0.5586 &  \textbf{0.52M} & \textbf{2.49$\times$} \\
    \bottomrule
    \end{tabular}}}
    \label{tab:quantitative_result}
\end{table*}

\begin{figure*}[t]
    \centering
    \subfigure[AdaRec-RetailRocket]{
    \centering
    \begin{minipage}[b]{0.9\textwidth}
    \includegraphics[width=0.9\textwidth]{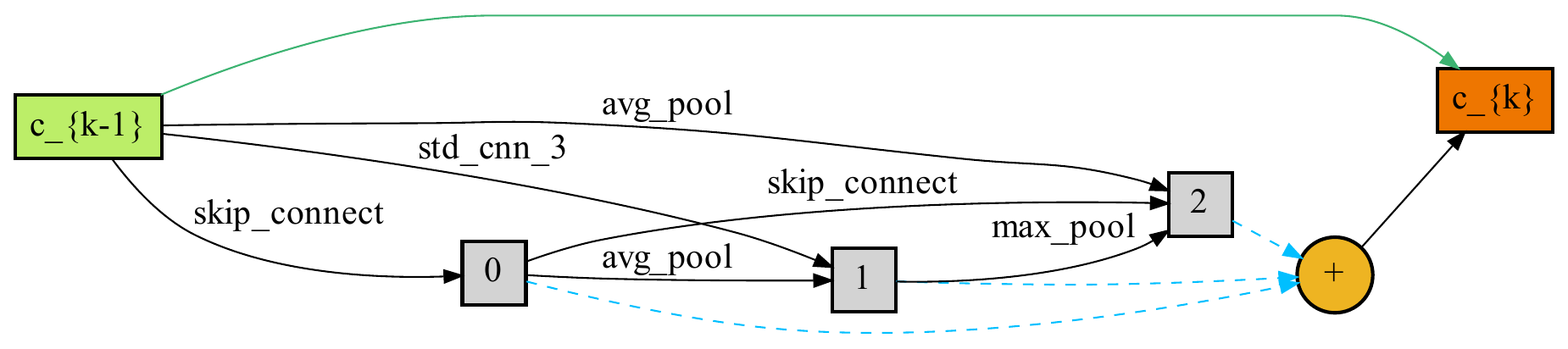}
    \end{minipage}
    }
    \subfigure[AdaRec-30Music]{
    \centering
    \begin{minipage}[b]{0.9\textwidth}
    \includegraphics[width=0.9\textwidth]{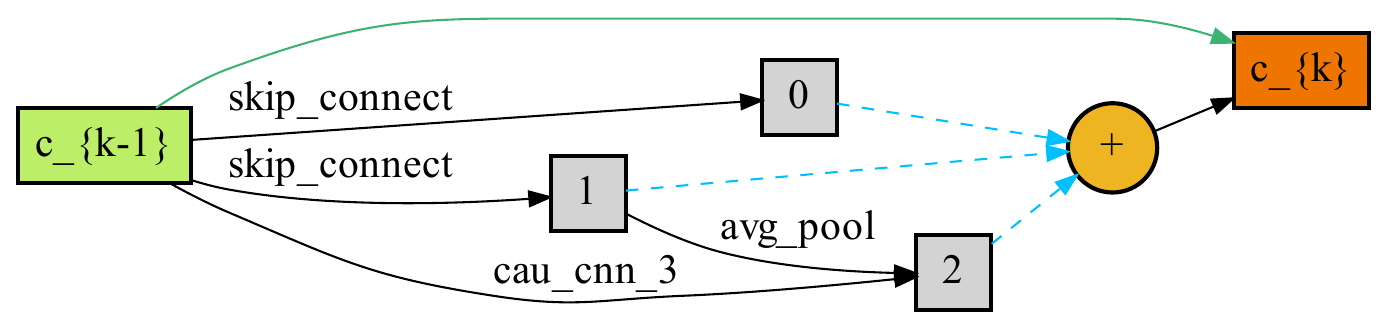}
    \end{minipage}
    }
    \subfigure[AdaRec-ML-2K]{
    \centering
    \begin{minipage}[b]{0.9\textwidth}
    \includegraphics[width=0.9\textwidth]{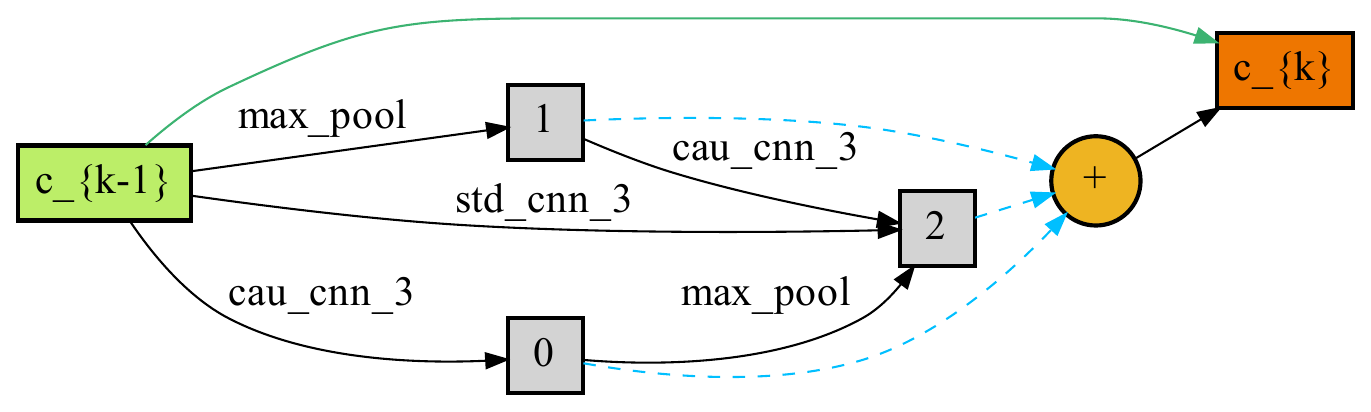}
    \end{minipage}
    }
    \caption{Visualization of the basic blocks of AdaRec on the three recommendation scenarios by using NextItNet as the teacher model. $\mathit{std\_cnn\_3}$ and   $\mathit{cau\_cnn\_3}$ represent the standard and causal convolutional layers with kernel size 3, respectively.}
    \label{fig:arch_visualization}
\end{figure*}

\subsection{Knowledge Distillation}

Specifically, we distill knowledge from the teacher model from three perspectives: the embedding layer, the hidden layers, and the prediction layer.

 \paragraph{Embedding Layer Distillation} The prediction accuracy of the sequential recommendation model, such as NextItNet, can be largely improved by increasing the embedding dimension~\cite{sun2020generic}.
Compressing item embedding matrices without sacrificing model performance is essential online inference speedup and parameter reduction. We define $\mathcal{L}_{\text {emb}}$ as the distillation loss of the embedding layer, where it 
is minimized by the mean squared error (MSE) between the teacher model and the student model:
\begin{equation}
  \mathcal{L}_{\text {emb}}=\operatorname{MSE}\left(\mathbf{E}^{T}, \mathbf{E}^{S} \mathbf{W}_{e}\right)  
\end{equation}
where $\mathbf{E}^{T}$ and $\mathbf{E}^{S}$ represent the item embedding matrices of teacher and student models, respectively. $\mathbf{W}_{e}$ is a learnable projection parameter.
\paragraph{Prediction Layers Distillation} The student model is encouraged to match the prediction ability of the teacher model by learning from the probability logits of the teacher model. We define $\mathcal{L}_{\text {pred}}$ using KL divergence~\cite{kullback1951information} as the distillation loss of the prediction layer:
\begin{equation}
\small
\mathcal{L}_{\text {pred}}=\mathit{KL}\left(\mathbf{z}^{T}, \mathbf{z}^{S} \right)
\end{equation}
where $\mathbf{z}^{T}$ and $\mathbf{z}^{S}$ are probability logits after passing through the softmax layer of the teacher \& student models, respectively. 

\paragraph{Hidden Layers Distillation} Since there are different numbers of hidden layers in teacher and  student, conventional one-to-one layer mapping algorithms cannot be applied.
Here, we employ the Earth Mover's Distance (EMD)~\cite{rubner2000earth} algorithm to encourage each student hidden layer to learn from multiple teacher layers adaptively. 
EMD measures the distance between the teacher and student networks as the minimum cumulative cost of knowledge transfer~\cite{rubner2000earth}. 

The key insight is to view network layers as distributions, and the desired transformation should make the two distributions (i.e., teacher and student layers) close. Formally, let $\mathbf{H}^{T}=\left\{\left(\mathbf{H}_{1}^{T}, w_{T_{1}}^{\mathbf{H}}\right), \ldots,\left(\mathbf{H}_{N}^{T}, w_{T_{N}}^{\mathbf{H}}\right)\right\}$ be the hidden layers of teacher model and $\mathbf{H}^{S}=\left\{\left(\mathbf{H}_{1}^{S}, w_{S_{1}}^{\mathbf{H}}\right), \ldots,\left(\mathbf{S}_{K}^{S}, w_{S_{K}}^{\mathbf{H}}\right)\right\}$ be the hidden layers of student model, where $\mathbf{H}_i^{T}$ and $\mathbf{H}_j^{S}$ represent the $i$-th and $j$-th hidden layer of the teacher and student models, $w_{T_{i}}^{\mathbf{H}}$ and $w_{S_{j}}^{\mathbf{H}}$ are corresponding layer weights, $N$ and $K$ represent the number of hidden layers in the teacher and student models, respectively. We define a ``ground'' distance matrix $\mathbf{D}^{\mathbf{H}}=\left[d_{i j}^{\mathbf{H}}\right]$, where $d_{i j}^{\mathbf{H}}$ represents the cost of transferring the knowledge of hidden states from $\mathbf{H}_i^{T}$ to $\mathbf{H}_j^{S}$. We adopt KL divergence to calculate the distance $d_{i j}^{\mathbf{H}}$:
\begin{equation}
 d_{i j}^{\mathrm{H}}=\operatorname{K L}\left(\mathbf{H}_{i}^{T}, \mathbf{H}_{j}^{S} \mathbf{W}_{h}\right)  
\end{equation}
where $\mathbf{W}_{h}$ is a learnable projection parameter.

Then, a mapping flow matrix $\mathbf{F}^{\mathbf{H}}=\left[f_{i j}^{\mathbf{H}}\right]$, with $f_{i j}^{\mathbf{H}}$ the mapping flow between $\mathbf{H}_i^{T}$ and $\mathbf{H}_j^{S}$, is learned by minimizing the cumulative cost required to transfer knowledge from $\mathbf{H}^{T}$ to $\mathbf{H}^{S}$:
\begin{equation}
\operatorname{WORK}\left(\mathbf{H}^{T}, \mathbf{H}^{S}, \mathbf{F}^{\mathbf{H}}\right)=\sum_{i=1}^{N} \sum_{j=1}^{K} f_{i j}^{\mathbf{H}} d_{i j}^{\mathbf{H}}
\end{equation}
subject to the following constraints:
\begin{equation}
\small
f_{i j}^{\mathrm{H}} \geq 0 \quad 1 \leq i \leq N, 1 \leq j \leq K 
\end{equation}
\begin{equation}
\small
\sum_{j=1}^{K} f_{i j}^{\mathbf{H}} \leq w_{T_{i}}^{\mathbf{H}} \quad 1 \leq i \leq N 
\end{equation}
\begin{equation}
\small
\sum_{i=1}^{N} f_{i j}^{\mathbf{H}} \leq w_{S_{j}}^{\mathbf{H}} \quad 1 \leq j \leq K 
\end{equation}
\begin{equation}
\small
\sum_{i=1}^{N} \sum_{j=1}^{K} f_{i j}^{\mathbf{H}}=\min \left(\sum_{i}^{N} w_{T_{i}}^{\mathbf{H}}, \sum_{j}^{K} w_{S_{j}}^{\mathbf{H}}\right)
\end{equation}

After solving the above optimization problem, we obtain the optimal mapping flow $\mathbf{F}^{\mathbf{H}}$. The Earth Mover’s Distance (EMD) can be defined as the work normalized by the total flow:
\begin{equation}
\operatorname{EMD}\left(\mathbf{H}^{S}, \mathbf{H}^{T}\right)=\frac{\sum_{i=1}^{N} \sum_{j=1}^{K} f_{i j}^{\mathbf{H}} d_{i j}^{\mathbf{H}}}{\sum_{i=1}^{N} \sum_{j=1}^{K} f_{i j}^{\mathbf{H}}}
\end{equation}

Finally, the hidden-layer distillation loss (termed as $\mathcal{L}_{\text {hidden}}$) can be defined by the EMD between $\mathbf{H}^{T}$ and $\mathbf{H}^{S}$:
\begin{equation}
\mathcal{L}_{\text {hidden}}=\operatorname{EMD}\left(\mathbf{H}^{S}, \mathbf{H}^{T}\right)
\end{equation}

By combining the above three distillation objectives ($\mathcal{L}_{\text{emb}}$, $\mathcal{L}_{\text{pred}}$, $\mathcal{L}_{\text{hidden}}$), we can unify the knowledge distillation loss $\mathcal{L}_{KD}$ between the teacher model and the student model:
\begin{equation}
\mathcal{L}_{\text {KD}} = \mathcal{L}_{\text {emb}} + \mathcal{L}_{\text {pred}} + \mathcal{L}_{\text {hidden}}
\label{loss_kd}
\end{equation}

\paragraph{Efficiency Constraint} We devise an efficiency constraint, which explicitly takes the efficiency of the student model into the main objective to achieve a trade-off between recommendation effectiveness and efficiency. Specifically, we define a cost-sensitive loss by considering both the parameter size and inference time:
\begin{equation}
\mathcal{L}_{E}=\sum_{o_{i, j} \in \alpha_{c}} SIZE\left(o_{i, j}\right)+F L O P s\left(o_{i, j}\right)
\label{loss_e}
\end{equation}
where $SIZE(\cdot)$ and $FLOPs(\cdot)$ are the normalized parameter size and the number of floating point operations (FLOPs) for each operation. The sum of FLOPs of searched operations can be used to approximate the actual inference time of the student model.

\begin{table*}[t]
\centering
\caption{Performance comparison on the three datasets for cross-scenario validation by using NextItNet as the teacher model.}
    \subtable[RetailRocket]{{\renewcommand{\arraystretch}{1.2}
    \begin{tabular}{L{2.5cm}|C{1.26cm}|C{1.26cm}|C{1.26cm}|C{1.26cm}|C{1.26cm}|C{1.26cm}}
    \toprule
    \multirow{2}{*}{\textbf{Architecture}} & \multicolumn{6}{c}{\textbf{RetailRocket}} \\ \cline{2-7}
    & MRR@5 & MRR@20 & HR@5 & HR@20 & NDCG@5 & NDCG@20 \\
    \midrule
    AdaRec-RetailRocket &\textbf{0.7345} & \textbf{0.7424} & \textbf{0.7964} & \textbf{0.8741} & \textbf{0.7500} & \textbf{0.7724}\\
    AdaRec-30Music & 0.7333 & 0.7412 & 0.7953 & 0.8743 & 0.7488 & 0.7155 \\ 
    AdaRec-ML-2K & 0.7283 & 0.7364 & 0.7926 & 0.8734 & 0.7444 & 0.7676 \\
    \bottomrule
    \end{tabular}}}
    \subtable[30Music]{{\renewcommand{\arraystretch}{1.2}
    \begin{tabular}{L{2.5cm}|C{1.26cm}|C{1.26cm}|C{1.26cm}|C{1.26cm}|C{1.26cm}|C{1.26cm}}
    \toprule
    \multirow{2}{*}{\textbf{Architecture}} & \multicolumn{6}{c}{\textbf{30Music}} \\ \cline{2-7}
    & MRR@5 & MRR@20 & HR@5 & HR@20 & NDCG@5 & NDCG@20 \\
    \midrule
    AdaRec-RetailRocket & 0.6164 & 0.6297 & 0.6956 & 0.8297 & 0.6361 & 0.6744 \\
    AdaRec-30Music & \textbf{0.6343} & \textbf{0.6473} & \textbf{0.7151} & \textbf{0.8452} & \textbf{0.6544} & \textbf{0.6917} \\ 
    AdaRec-ML-2K & 0.6248 & 0.6379 & 0.7056 & 0.8378 & 0.6449 & 0.6827 \\
    \bottomrule
    \end{tabular}}}
    \subtable[ML-2K]{{\renewcommand{\arraystretch}{1.2}
    \begin{tabular}{L{2.5cm}|C{1.26cm}|C{1.26cm}|C{1.26cm}|C{1.26cm}|C{1.26cm}|C{1.26cm}}
    \toprule
    \multirow{2}{*}{\textbf{Architecture}} & \multicolumn{6}{c}{\textbf{ML-2K}} \\ \cline{2-7}
    & MRR@5 & MRR@20 & HR@5 & HR@20 & NDCG@5 & NDCG@20 \\
    \midrule
    AdaRec-RetailRocket & 0.3248 & 0.3548 & 0.5137 & 0.7839 & 0.3717 & 0.4559 \\
    AdaRec-30Music & 0.3969 & 0.4207 & 0.5951 & 0.8357 & 0.4520 & 0.5135 \\ 
    AdaRec-ML-2K & \textbf{0.4489}  & \textbf{0.4732} & \textbf{0.6519} & 0.8825 & \textbf{0.4995} & \textbf{0.5670} \\
    \bottomrule
    \end{tabular}}}
    \label{tab:cross_validation}
\end{table*}

\subsection{Training Procedure}
\label{loss}
Follow the common practice, we first pre-train the teacher model and then search for the student structure under the supervision of the pre-trained teacher model. When searching the student architecture, we combine the knowledge distillation loss $\mathcal{L}_{\text{KD}}$  and the 
cost-sensitive loss $\mathcal{L}_{\text{E}}$.
Besides, we also need to incorporate the cross-entropy loss ($\mathcal{L}_{CE}$) w.r.t ground-truth labels from the training data to assist the searching process, which is defined as:
\begin{equation}
{\mathcal{L}_{CE}} = - \sum_{X^{u}\in \mathbf{X}} {p(x_{t+1}^u)} \log {p(\hat{x}_{t+1}^u)}
\label{loss_ce}
\end{equation}
where $\mathbf{X}$ represents the whole user-item interaction sequences in the training data, $p(x_{t+1}^u)$ is the ground truth distribution for next item prediction and $p(\hat{x}_{t+1}^u)$ is the prediction distribution of the searched student model. 

The overall loss function is defined as follows:
\begin{equation}
\label{loss_all}
\mathcal{L}=(1-\gamma) \mathcal{L}_{C E} + \gamma \mathcal{L}_{K D} + \beta \mathcal{L}_{E}
\end{equation}
where $\gamma$ \& $\beta$ are hyper-parameters that balance these loss functions.

After finishing the joint search of student structure and knowledge transfer under the guidance of the pretrained teacher model, we can derive an effective, efficient and adaptive architecture as the compressed sequential model by stacking the searched block structures.

\paragraph{Differentiable Neural Architecture Searching} 
Directly optimizing the objective function in Eq. (\ref{loss_all}) by brute-force enumeration of all candidate operations is impossible due to the huge search space with combinatorial operations. 
To resolve such an issue, we model the search operation $o_{i,j}$ as discrete variables (one-hot variables) that obey the discrete probability distributions $P_{o}=\left[\theta_{1}^{o}, \ldots, \theta_{|\mathcal{O}|}^{o}\right]$. 
Then, we use a Gumbel-Softmax distribution \cite{maddison2016concrete} to relax the categorical samples into continuous vectors $y^{o} \in R^{\mathcal{O}}$ as: 
\begin{equation}
\mathbf{y}_{i}^{o}=\frac{\exp \left[\left(\log \left(\theta_{i}^{o}\right)+g_{i}\right) / \tau\right]}{\sum_{j=1}^{|\mathcal{O}|} \exp \left[\left(\log \left(\theta_{j}^{o}\right)+g_{j}\right) / \tau\right]}
\end{equation}
where $g_i$ is a random noise drawn from Gumbel(0, 1) distribution, $\tau$ is a temperature coefficient to control the discreteness of the output vectors $\mathbf{y}^{o}$. In this way, we can optimize the objectives $\mathcal{L}_{KD}$ and $\mathcal{L}_{E}$ directly using gradient-based optimizers by using the discrete variable $argmax(\mathbf{y}^{o})$ in the forward pass and using the continuous vector $\mathbf{y}^{o}$ in the back-propagation stage.

\section{Experimental Setup}
\subsection{Experimental Datasets}
We conduct extensive experiments on three real-world SRS datasets from three different domains (scenes): RetailRocket from the E-commerce domain, 30Music from the music domain \cite{ludewig2019performance}, and MovieLens-2K from the movie domain~\cite{Cantador:RecSys2011}. The statistics of them are provided in Table~\ref{tab:data_statistic}.
\begin{itemize}[leftmargin=*]
    \item \textbf{RetailRocket}\footnote{\small https://retailrocket.net/}   contains user purchasing and clicking behaviors. We set the maximum length of each  sequence $t$ to 10 so as to investigate the recommendation performance with short-range sequences. We split the sequences longer than $t$ into multiple sub-sequences, while the ones shorter than $t$ are padded with zero in the beginning of each sequence, similar to \cite{yuan2019simple}. 
    \item \textbf{30Music} is a collection of listening and playlists data retrieved from Internet radio stations through Last.fm API\footnote{https://www.last.fm/}. We process it as a middle-range sequential dataset by extracting the latest 20 actions per user.
    \item \textbf{MovieLens-2K}\footnote{\small https://grouplens.org/datasets/hetrec-2011/} (denoted as ML-2K) is a benchmark dataset for both standard collaborative filtering and sequential recommendations. We set $t=50$  to evaluate the performance with long-range sequences.
\end{itemize}

\subsection{Baselines and Evaluation Metrics}
To verify the effectiveness and efficiency of AdaRec, we  compare it with its teacher model includingNextItNet~\cite{yuan2019simple}, SASRec~\cite{kang2018self} and BERT4Rec~\cite{sun2019bert4rec}  which have been described in Section~\ref{teacher}. In addition, we have also compare it with GRU4Rec~\cite{hidasi2015session} and Caser~\cite{tang2018personalized} for reference given that the two models are recognized as two most typical sequential recommendation baselines. 
Following \cite{yuan2020future}, we train Caser using the data augmentation method and train GRU4Rec based on the auto-regressive method.
To evaluate recommendation accuracy, we adopt three popular top-$N$ ranking metrics, including  MRR@$N$ (Mean Reciprocal Rank), HR@$N$ (Hit Ratio) and and NDCG@$N$ (Normalized Discounted Cumulative Gain)~\cite{yuan2020one}. Here $N$ is set to 5 and 20 for comparison. 
To evaluate the computational efficiency of AdaRec, we compare its model size and inference speedup with the teacher models.

\begin{table*}[t]
    \centering
    \caption{Performance comparison on the three datasets for loss ablation studies by using NextItNet as the teacher model.}
    \subtable[RetailRocket]{{\renewcommand{\arraystretch}{1.2}
    \begin{tabular}{L{2.5cm}|C{1.26cm}|C{1.26cm}|C{1.26cm}|C{1.26cm}|C{1.26cm}|C{1.26cm}}
    \toprule
    \multirow{2}{*}{\textbf{Model}} & \multicolumn{6}{c}{\textbf{RetailRocket}} \\ \cline{2-7}
    & MRR@5 & MRR@20 & HR@5 & HR@20 & NDCG@5 & NDCG@20 \\
    \midrule
    AdaRec (All) &\textbf{0.7345} & \textbf{0.7424} & \textbf{0.7964} & \textbf{0.8741} & \textbf{0.7500} & \textbf{0.7724} \\
    \multicolumn{1}{l|}{w/o $\mathcal{L}_{KD (emb)}$}  & 0.7239 & 0.7321 & 0.7886 &  0.8708 & 0.7401 & 0.7636  \\ 
    \multicolumn{1}{l|}{w/o $\mathcal{L}_{KD (pred)}$}  & 0.6898 & 0.6992 & 0.7583 & 0.8517 & 0.7070 & 0.7337 \\ 
    \multicolumn{1}{l|}{w/o $\mathcal{L}_{KD (hidden)}$}  & 0.7142 & 0.7227 & 0.7806 & 0.8647 & 0.7309 & 0.7551 \\ 
    \multicolumn{1}{l|}{w/o $\mathcal{L}_{CE}$} & 0.7115 & 0.7198 & 0.7804  & 0.8637 & 0.7288 & 0.7526 \\
    \bottomrule
    \end{tabular}}}
    \subtable[30Music]{{\renewcommand{\arraystretch}{1.2}
    \begin{tabular}{L{2.5cm}|C{1.26cm}|C{1.26cm}|C{1.26cm}|C{1.26cm}|C{1.26cm}|C{1.26cm}}
    \toprule
    \multirow{2}{*}{\textbf{Model}} & \multicolumn{6}{c}{\textbf{30Music}} \\ \cline{2-7}
    & MRR@5 & MRR@20 & HR@5 & HR@20 & NDCG@5 & NDCG@20 \\
    \midrule
    AdaRec (All) & \textbf{0.6343} & \textbf{0.6473} & \textbf{0.7151} & \textbf{0.8452} & \textbf{0.6544} & \textbf{0.6917} \\
    \multicolumn{1}{l|}{w/o $\mathcal{L}_{KD (emb)}$}  & 0.6193 & 0.6326 & 0.7055 & 0.8386 & 0.6408 &  0.6789 \\ 
    \multicolumn{1}{l|}{w/o $\mathcal{L}_{KD (pred)}$}  & 0.5512 & 0.5646 & 0.6218 & 0.7604 & 0.5687 & 0.6080 \\ 
    \multicolumn{1}{l|}{w/o $\mathcal{L}_{KD (hidden)}$}  & 0.6186 & 0.6319 & 0.7045 & 0.8378 & 0.6400 & 0.6782 \\ 
    \multicolumn{1}{l|}{w/o $\mathcal{L}_{CE}$} & 0.5976 & 0.6113 & 0.6899  & 0.8261 & 0.6206 & 0.6597 \\
    \bottomrule
    \end{tabular}}}
    \subtable[ML-2K]{{\renewcommand{\arraystretch}{1.2}
    \begin{tabular}{L{2.5cm}|C{1.26cm}|C{1.26cm}|C{1.26cm}|C{1.26cm}|C{1.26cm}|C{1.26cm}}
    \toprule
    \multirow{2}{*}{\textbf{Model}} & \multicolumn{6}{c}{\textbf{ML-2K}} \\ \cline{2-7}
    & MRR@5 & MRR@20 & HR@5 & HR@20 & NDCG@5 & NDCG@20 \\
    \midrule
    AdaRec (All)& \textbf{0.4489}  & \textbf{0.4732} & \textbf{0.6519} & \textbf{0.8825} & \textbf{0.4995} & \textbf{0.5670} \\
    \multicolumn{1}{l|}{w/o $\mathcal{L}_{KD (emb)}$}  & 0.4401 & 0.4650 & 0.6429 & 0.8786 & 0.4906 & 0.5595  \\ 
    \multicolumn{1}{l|}{w/o $\mathcal{L}_{KD (pred)}$}  & 0.2949 & 0.3254 & 0.4729 & 0.7729 & 0.3390 & 0.4255 \\ 
    \multicolumn{1}{l|}{w/o $\mathcal{L}_{KD (hidden)}$}  & 0.4351 & 0.4599 & 0.6430 & 0.8805 & 0.4868 & 0.5561 \\ 
    \multicolumn{1}{l|}{w/o $\mathcal{L}_{CE}$} & 0.4391 & 0.4642 & 0.6407  & 0.8794 & 0.4893 & 0.5590 \\
    \bottomrule
    \end{tabular}}}
    \label{tab:loss_reduction}
\end{table*}

\subsection{Implementation Details}
We divide the user-item interaction sequence $X^u=[x_{1:t}^u]$ for each user $u$ into $X_{train}^u=[x_{1:t-2}^u]$ for training, $x_{t-1}^u$ for validation and $x_{t}^u$ for testing, following~\cite{kang2018self}.
For the teacher model NextItNet, we set the embedding dimension $d$ to be 256, and use dilation factors of $8\times\{1, 2, 4, 8\}$ (32 layers or 16 residual blocks). For both SASRec and BERT4Rec, we set  $d$ to be 128 given that a larger  $d$ hurts their performance because of overfitting. We use 8 self-attention blocks with four heads for SASRec and BERT4Rec according to its accuracy in the validation set.
When searching the architecture for the student model, we set $d$ to one quarter of its teacher's embedding dimension (i.e., $d=64$ for NextItNet and $d=32$ for SASRec and BERT4Rec), $\gamma=0.5$, $\beta=8$, inner nodes $M=3$ and student blocks $K=4$. 
For training AdaRec, we employ AdamW~\cite{loshchilov2018fixing} to optimize the parameters (e.g., embedding matrix and searched operations) with learning rate $\eta = 5e-3$ and weight decay of $5e-4$, and architecture distribution $P_{o}$ with learning rate $\eta = 2e-5$ and weight decay of $1e-4$.
All the experiments are implemented in PyTorch and trained on a single TITAN RTX GPU.

\section{Experimental Results}
\subsection{Overall Results}
Table \ref{tab:quantitative_result} reports the performance (i.e., MRR@$N$, HR@$N$ and NDCG@$N$ ($N$ is set to 5 and 20), parameter size and inference speedup) of AdaRec and  baseline models on the three datasets. From the results, we can make the following observations. First, we observe that NextItNet, SASRec and BERT4Rec outperform GRU4Rec and Caser with substantial improvements in terms of recommendation accuracy among the three datasets, which is consistent with the previous work~\cite{yuan2019simple,kang2018self,sun2019bert4rec}. Second, AdaRec with NextItNet, SASRec and BERT4Rec as teacher models attain competitive or 
better recommendation accuracy than their teachers, although we do not expect AdaRec beats its teacher model in accuracy. For example, on RetailRocket and 30Music, AdaRec with NextItNet as the teacher model obtains 2.9$\%$ and 3.2$\%$ improvements over its large teacher model in terms of MRR@5. Importantly, AdaRec requires much fewer parameters and achieves notable inference speedup relative to its teachers. In addition, compared to the standard KD method~\cite{hinton2015distilling} with equivalent model size,  AdaRec with NAS techniques performs substantially better with higher inference speedup. 

\subsection{Cross-Scene Evaluation}
In this section, we investigate the scene-adaptivity of AdaRec with different recommendation scenarios.
We apply the searched student architecture from one recommendation scenario to other  scenarios. For example, we denote the searched student architecture for RetailRocket (i.e., E-commerce domain) with  NextItNet as the teacher model as AdaRec-RetailRocket, and apply it to 30Music (i.e., music domain) and ML-2K (i.e., movie domain). For such cross-scenario validation, we randomly initialize the weights of each searched student structure and re-train it using corresponding training data and the same teacher model to ensure a fair comparison. The results are summarized in Table~\ref{tab:cross_validation}, where we omit results using SASRec and BERT4Rec as teacher models due to similar behaviors. 
As  clearly demonstrated along the diagonal line of Table~\ref{tab:cross_validation}, we can draw  that AdaRec achieves the best performance on their original recommendation scenarios in contrast to other scenarios. This is, AdaRec is scene-adaptive since the searched student network only guarantees its optimal performance on a specific recommendation scenario.

\begin{figure*}[t]
\centering
\subfigure[\small RetailRocket]{
\includegraphics[width=0.9\columnwidth]{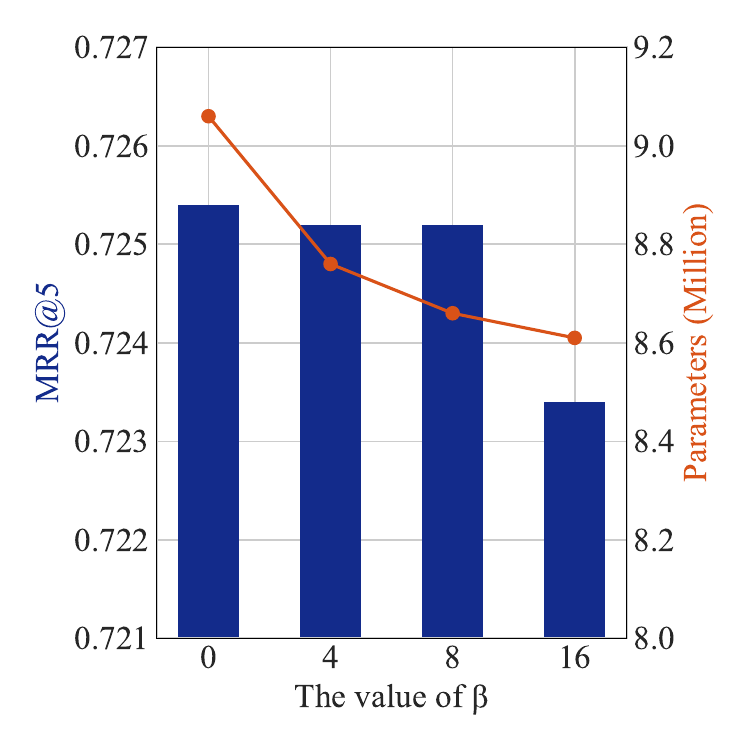}}
\subfigure[\small 30Music]{
\includegraphics[width=0.9\columnwidth]{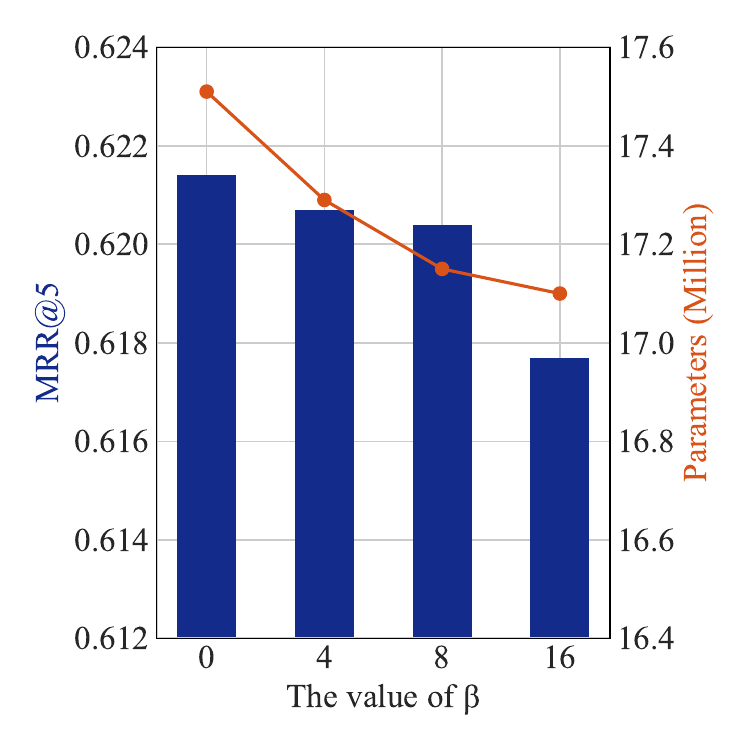}}
\caption{Performance comparison on RetailRocket and 30Music for varying coefficient $\beta$ of the cost-sensitive loss ($\mathcal{L}_{E}$) by using NextItNet as the teacher model.}
\label{fig:beta}
\end{figure*}

\subsection{Architecture Visualization}
To better understand the basic blocks of the searched student architectures, we visualize them on the three recommendation scenarios in Figure~\ref{fig:arch_visualization}. For space reason, we still only show AdaRec with NextItNet as the teacher model. By comparing the searched structures for different recommendation scenarios, we can find that AdaRec for RetailRocket (from E-commerce domain) and 30Music (from music domain) are relatively lightweight, since 
fewer convolution operations (i.e., $\mathit{std\_cnn\_3}$ for RetailRocket and $\mathit{cau\_cnn\_3}$ for 30Music) are used. This is likely because the two datasets have short-range sequential dependencies. On the contrary, a more complicated student structure with  diverse convolution operations(i.e., $\mathit{std\_cnn\_3}$ and $\mathit{cau\_cnn\_3}$)  is learned for ML-2K so as to model the long-range dependencies. 
The above results well back up our claim that the proposed AdaRec is able to search adaptive student structures for different recommendation scenarios.
 
\subsection{Ablation Studies}
As described before, the loss $\mathcal{L}$ of AdaRec consists of three parts: the target-oriented KD loss $\mathcal{L}_{KD}$, the cost-sensitive loss $\mathcal{L}_{E}$ and the standard cross-entropy loss $\mathcal{L}_{CE}$. First, we evaluate the effects of $\mathcal{L}_{KD}$ and $\mathcal{L}_{CE}$ by removing each of them independently, as reported in Table~\ref{tab:loss_reduction}. Clearly, we find that AdaRec without each of the two losses  yields  sub-optimal recommendation accuracy on all three datasets. 
Besides, it also shows that combining distillation losses on the embedding layer  $\mathcal{L}_{\text {emb}}$, prediction layer   $\mathcal{L}_{\text {pred}}$ and hidden layers  $\mathcal{L}_{\text {hidden}}$
together produces the best results.

In addition, we verify the effect of the cost-sensitive loss $\mathcal{L}_{E}$ by varying $\beta$, including the default case $\beta=8$, without constraint $\beta=0$, weak  constraint $\beta=4$ and strong constraint $\beta=16$. The model performance and corresponding model size are illustrated in Figure~\ref{fig:beta}. From the results we can see that no  constraint or a small value of $\beta$ lead to an increased model size; meanwhile, an aggressive $\beta$ results in a smaller model size but degraded model accuracy on the other hand. An appropriate constraint ($\beta=8$) achieves the superior trade-off between the model effectiveness and  efficiency.

\section{Conclusion}
In this paper, we present a novel sequential recommendation  knowledge distillation (KD) framework AdaRec based on the differentiable Neural Architecture Search (NAS). AdaRec  compresses  knowledge  of  large and deep sequential recommendation models into  a compact student  model  adaptively  with the recommendation scene. 
To the best of our knowledge, AdaRec is the first to propose combining knowledge distillation and neural architecture search in the SRS tasks so as to adaptively compress the deep sequential recommendation models according to different recommendation scenes. 
Besides, we devise the Earth Mover's Distance
(EMD) based KD method for effective transfer of deep hidden layers and introduce a cost-sensitive constraint to achieve the trade-off between effectiveness and efficiency of SRS tasks. Extensive experiments on three real-world recommendation datasets from different scenarios demonstrate that AdaRec achieves considerably better accuracy compared to the standard KD baseline and comparable accuracy with its teacher model while  accelerating inference time and reducing the computational workload largely.

\section*{Acknowledgement}
Min Yang was partially supported by the National Natural Science Foundation of China (NSFC) (No. 61906185), Youth Innovation Promotion Association of CAS China (No. 2020357), Shenzhen Science and Technology Innovation Program (No. KQTD20190929172835662), 
Shenzhen Basic Research Foundation (No. JCYJ20200109113441941). 
\ifCLASSOPTIONcaptionsoff
  \newpage
\fi

\bibliographystyle{IEEEtran}
\bibliography{ref.bib}

\end{document}